# First-principles simulation of capacitive charging of graphene and implications for supercapacitor design

Maxwell D. Radin[1], Tadashi Ogitsu[2], Minoru Otani[3], Juergen Biener[2], Brandon C. Wood[2]

[1]*Department of Physics, University of Michigan, Ann Arbor, MI 48105*

[2]*Materials Science Division, Lawrence Livermore National Laboratory, Livermore, CA 94550*

[3]*National Institute of Advanced Industrial Science and Technology, Tsukuba, Japan*

**Abstract:** Supercapacitors store energy via the formation of an electric double layer, which generates a strong electric field at the electrode-electrolyte interface. Unlike conventional metallic electrodes, graphene-derived materials suffer from a low electronic density of states (i.e., quantum capacitance), which limits their ability to redistribute charge and efficiently screen this field. To explore these effects, we introduce a first-principles approach based on the effective screening medium framework, which is used to directly simulate the charge storage behavior of single- and multi-layered graphene in a way that more closely approximates operating devices. We demonstrate that the presence of the interfacial field significantly alters the capacitance in electrodes thinner than a few graphene layers, deriving in large part from intrinsic space-charge screening limitations. The capacitance is also found to be highly sensitive to the gap between the electrode and the solvent (contact layer), which offers possibilities for tuning the interfacial capacitance of the electrode by proper engineering of the electrolyte. Our results offer an alternative interpretation of discrepancies between experimental measurements and fixed-band models, and provide specific implications for improving graphene-based devices.

Graphene-derived supercapacitors are a promising technology for electrochemical energy storage due to their high specific areas, high conductivities, structural and chemical tunability, and use of environmentally benign materials [1–5]. However, the measured capacities of these devices are severely limited by the number of electronic states available near the Fermi level in the solid-state electrode, sometimes called the quantum capacitance [6–10]. Understanding the response of the electrode's electronic structure to applied potentials is therefore essential for assessing and optimizing supercapacitor performance.

In supercapacitors, an "electrical double layer" (EDL) is created at the interface with the electrode as the ionic electrolyte solution responds dynamically to charge accumulation or depletion. The presence of the EDL induces a strong electric field at the electrode-electrolyte interface, which alters the charge distribution within the electrode and creates a space-charge layer. For conventional metallic electrodes, efficient screening minimizes the impact of this redistribution; however, this is not the case for graphene-derived systems, where screening is poor due to a low electronic density of states (DOS). Prior theoretical studies [9,11,12] on the quantum capacitance of single-layer graphene (SLG) have tended to rely upon the fixed-band approximation (FBA), in which electrode charging or discharging shifts the overall Fermi level as bands fill or deplete. However, the FBA ignores the effect of the interfacial field and instead assumes a homogenous distribution of net charge carriers.

In this letter, we introduce a new model for simulating capacitive charging from first principles, from which the quantum capacitance can be directly extracted. Our approach relies on the effective screening medium (ESM) method [13] to go beyond the FBA and explicitly account for the effects of the EDL-induced interfacial electric field, more closely approximating conditions experienced during device operation. We demonstrate that for graphene electrodes, charge redistribution at the interface significantly impacts the observed quantum capacitance, and that the nature of the variations offers valuable insights for tuning capacitance in real devices.

The capacitance associated with the electrode-electrolyte interface can be thought of as two capacitances in series, one associated with the liquid electrolyte and the second with the solid electrode (i.e., split by a Gibbs dividing surface). While the accumulation of charge in the first is determined by ion concentration and thermal reorganization, in the second it is typically limited by quantum effects, i.e., the "quantum" capacitance. Formally, the area-specific differential quantum capacitance is given by $C_q = d\sigma/dV_q$, where $\sigma$ is the surface charge density on the electrode and $V_q = -\mu/e$ measures the chemical potential of electrons $\mu$ relative to the energy of an electron at the electrostatic potential of the dividing surface (plus an arbitrary constant). Here $e$ is the elementary charge. The quantum capacitance of graphene electrodes has conventionally been estimated via the FBA, $C_q^{FBA} = e^2 n(\mu^{FBA}, N_0)$ [9,11,12]. Here $n(E,N)$ is the areal density of states (DOS) at energy $E$ when the system has a total of $N = N_0 - \sigma/e$ electrons per unit area, and $\mu^{FBA}$ is implicitly defined by $\int_{-\infty}^{\mu^{FBA}} n(E, N_0) dE = N$.

The FBA neglects some physical phenomena, including the accumulation of excess charge near surfaces in accordance with Gauss's law. Such phenomena are accounted for in the more general relation

$$C_q = e^2 \frac{n(\mu_e, N)}{1 - \int_{-\infty}^{\mu_e} \frac{\delta n(E,N)}{\delta N} dE}. \tag{0}$$

(See Supplemental Material for derivation.) Eq. 1 reduces to the FBA when $\int_{-\infty}^{\mu_e} \delta n(E,N)/\delta N \, dE$ is small, i.e., when the DOS is unaffected by excess charge. We will show that although qualitative features are reproduced, the FBA provides an incomplete quantitative description once we consider electronic screening limitations and charge redistribution within graphene electrodes in the presence of the EDL.

In this work, we directly compute the response of $V_q$ to $\sigma$ within the ESM framework, [13] yielding a more realistic view of charge and discharge. The ESM method inserts an infinite-dielectric medium at the boundary of the periodic simulation cell. Buildup of a mirror charge within this medium generates an interfacial electric field as the electrode is charged or discharged, mimicking the actual response of a supercapacitor electrode in contact with an ideal-screening "perfect" electrolyte.

Plane-wave pseudopotential density functional theory (DFT) calculations were performed using the van der Waals density functional (vdW-DF) [14–16] with a revised Perdew-Burke-Ernzerhof (revPBE) exchange reference, [17,18] as implemented in Quantum Espresso [19]. Full geometry optimizations were done using a 24 × 24 × 1 k-point mesh within the primitive cell, and forces were converged to within $10^{-5}$ Ry/Bohr. A finer k-point grid of 96 × 96 × 1 was used for the capacitance calculations. A vacuum spacing of > 9 Å separated periodic images.

The capacitance of single- and multi-layered graphene electrodes is determined within two distinct configurations, shown schematically in Fig. 1. In the first (Fig. 1a), a perfect-screening medium is placed parallel to the surface at one of the cell edges; this configuration mimics an electrode with one side in contact with an electrolyte (one-sided charging). In the second (Fig. 1b), two perfect-screening media are placed at opposite cell edges, which mimics an electrode with both sides in contact with an electrolyte (two-sided charging). These two representative configurations were chosen to span a variety of device setups and graphene-derived materials [1–5,20].

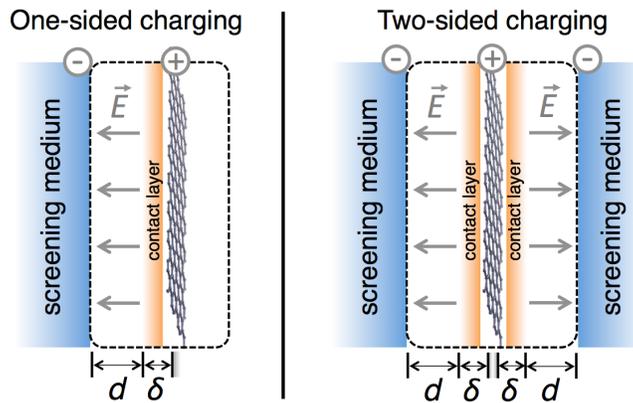

*FIG. 1. Simulation setup for one- and two-sided charging within the ESM method (shown here for positive bias).*

One could compute the quantum capacitance by placing the cell boundary at the position of the dividing surface, which lies a distance $\delta$ from the surface carbon nuclei. However, this would require a different set of DFT calculations to be performed for different positions of the dividing surface. Instead, we place the cell boundary a fixed distance away from the slab and map the system on to an equivalent circuit of two series capacitances: one associated with the differential quantum capacitance of the electrode $C_q$, and the other with the capacitance $C_v$ due to the potential drop across the vacuum between the dividing surface and screening medium [8]:

$$\frac{1}{C} = \frac{1}{C_q} + \frac{1}{C_v}. \tag{0}$$

For two-sided charging, the last term on the right-hand side of Eq. 2 is doubled because there are contributions from two gap regions. The total capacitance of the simulation cell $C$ was determined by numerically differentiating the $\sigma$-$V$ relationship obtained from a series of single-point calculations at different values of $\sigma$. $V$ was calculated as the Fermi level relative to the energy of a free electron at the electrostatic potential in the screening region(s). We model $C_v$ in Eq. 2 as a parallel plate capacitor, $C_v = \varepsilon_0/d$, where $d$ is the distance from the dividing surface to the screening medium. The voltage drop $V_q$ associated with the electrode is given by

$$V_q(\sigma) - V_q(0) = \int_0^\sigma \frac{1}{C_q(\sigma')} d\sigma', \tag{0}$$

where $V_q(0)$ corresponds to the potential of zero charge.

The value of $\delta$ is in principle arbitrary and reflects the partitioning of the voltage drop between the electrode and electrolyte. However, it is convenient to choose the position of the Gibbs dividing surface as the position where the screening of electric fields by the electrolyte begins; then the quantum capacitance represents the capacitance that would be measured if the electrolyte were a perfect screening medium. Thus we envision $\delta$ as representing the inner boundary of the electrolyte (see Fig. 1). For now, we adopt the carbon van der Waals radius of $r_w$ = 1.70 Å for $\delta$ [21]; this choice is motivated by the fact that the van der Waals radius gives a good description of cavities in continuum solvation models, as it represents the approximate distance to electrons in an electrolyte [22]. We will return a detailed discussion of the significance of $\delta$ later.

Applying this approach to single- and multi-layered graphene gives the differential quantum capacitance profiles shown in Fig. 2. We initially focus our discussion on SLG (red curves). It is immediately evident that the magnitude of the slope of the capacitance with voltage predicted by ESM (based on $\delta$ = 1.70 Å) is much smaller than that predicted by the FBA (dashed line). This means that the presence of the interfacial electric field reduces the effective capacitance of the graphene electrode. (The minimum in the capacitance corresponds to the Dirac point, at which the DOS for SLG is in principle zero; in practice, the electronic smearing used in DFT calculations result in a non-zero DOS.)

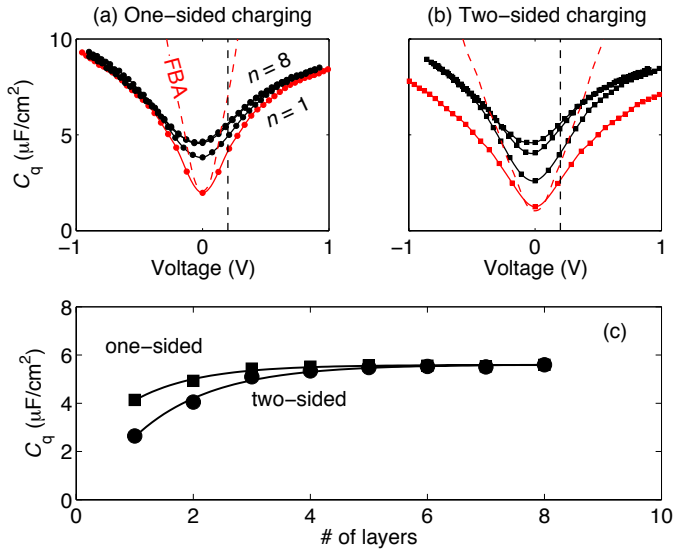

FIG. 2. *Dependence of calculated area-specific differential quantum capacitance on the number of graphene layers* n *for (a) one-sided and (b) two-sided charging, for slabs 1 (in red), 2, 4 and 8 layers thick. The dashed red line shows the FBA-derived capacitance. (c) Comparison of values at*

*a bias of +0.2 V (dashed vertical line in (a) and (b)). The solid lines show a fit to an exponential form.*

Experimental measurements of the quantum capacitance of single-layer graphene [9,11,12,23–25] exhibit significant variability. For example, estimates of the capacitance at a bias of 0.2 V relative to the potential of minimum capacitance range from ~2 to ~10 $\mu F/cm^2$ [11,12]. Nevertheless, the available results generally exhibit a shallower-than-expected capacitance dependence with respect to band-filling models, with slopes in good agreement with Fig. 2. Although experimental complexities (such as impurities [11] or local potential fluctuations [23]) have been invoked to reconcile the measured deviations from expected behavior, our results demonstrate that these same patterns are intrinsic to pristine graphene once the effect of the interfacial electric field is taken into account. Our data may also explain why Fermi velocities of graphene (proportional to the inverse slope of the dependence in Fig. 2) tend to be ~10% higher when obtained from best-fit models to electrochemical data compared direct measurements by other methods where no interfacial field is present [11,23,26,27].

Charge redistribution influences the relative capacitances of one- and two-sided charging of SLG. The FBA predicts one-sided charging to have twice the capacitance of one-sided charging. In contrast, Fig. 2 illustrates that the ratio of the ESM capacitance of SLG during one-sided charging to that during two-sided charging lies in the range 1.2-1.6 over the voltage range considered. The reduction in one-sided capacitance relative to two-sided capacitance can be attributed to the redistribution of excess charge towards the side of the graphene facing the EDL, an effect not accounted for in the FBA.

Such space-charge effects are likely responsible for the anomalously low area-specific capacitance observed in high-surface area carbon electrodes [1,6] as the pore wall thickness may be smaller than the screening length [8]. To probe this effect, we examined the dependence of capacitance on electrode thickness. Results for graphitic *AB* slabs with the number of layers *n* going up to eight are shown alongside the SLG values in Fig. 2. As the thickness increases, space-charge screening limitations become less severe. Accordingly, the capacitances for both one- and two-sided charging increase and ultimately converge. The magnitude of the space-charge effect is voltage dependent, and is most evident at potentials near the capacitance minimum, where the DOS is lowest (the notable exception is SLG under two-sided charging, which exhibits the largest space-charge effect of all). For instance, at +0.2 V (Fig. 2c), we find that for one-sided charging, the capacitance plateaus at ~3 layers, beyond which the electric field is fully screened. For two-sided charging, slightly thicker slabs (~5 layers) are required to reach the maximum capacitance because both sides of the electrode must now screen the field. We can fit the curves in Fig. 2c to an exponential of the form $C_q(n) = C_\infty - \Delta C \exp[-\kappa(n-1)]$ to estimate a screening length $\kappa^{-1}$ of 1.1 and 1.3 layers for one- and two-sided charging, respectively (~1.8 and 2.2 Å). These values are in reasonable agreement with our prior analysis based on polarization effects [10].

A close inspection of Figs. 2a and 2b shows that in addition to an increase in capacitance with the number of layers, there is also a small shift in the position of the minimum towards lower potentials. The potential of minimum capacitance decreases by ~20 mV as the number of layers goes from one to eight. This is consistent with the fact that in graphite, the Fermi level lies not at the minimum of the DOS, but slightly below the minimum [28]. It also agrees with electrochemical experiments that have found the potential of minimum capacitance of glassy carbon to be ~20 mV lower than that of an activated carbon aerogel, whose building blocks are closer to SLG [7].

Next, we return to a discussion of our chosen definition for $\delta$. Recent studies have reported a region of low electron density between the electrode and the Stern layer of the electrolyte [29,30]. Ando *et al.* termed this region the "contact layer", which they showed exists due to Pauli repulsion at the interface (in the absence of specific adsorption) [29]. Here, we associate the contact layer with the electrode capacitance, and therefore include it in our definition of $\delta$ (see Fig. 1); however, we emphasize that in reality, it is a purely interfacial characteristic that derives from the interaction between the electrode and the electrolyte.

The low electron density in the contact layer leads to poor screening and a concomitant voltage drop that influences the capacitance. The sensitivity of the calculated capacitance to $\delta$ is shown in Fig. 3, with the effects most pronounced at high voltage magnitudes and for small values of $\delta$. To interpret this result, consider that physically, the contact layer thickness is related to the approach distance of the electrolyte to the electrode, and more particularly to the distance at which the electronic density associated with the electrolyte atoms becomes appreciable. This distance can depend on several factors, including the interaction strength between the electrode and electrolyte (e.g., surface hydrophilicity/hydrophobicity), as well as the voltage. Accordingly, Fig. 3 asserts that the identity of the electrolyte could have a significant effect on the interfacial capacitance of the electrode, even in the limit of a short Debye length in the liquid. This finding is directly analogous to the recent discovery of strong capacitance changes due to the formation of a low-dielectric interface in nanoscale capacitors, also revealed by first-principles calculations [31].

In addition, local variations in the electrolyte approach distance will broaden our single value for $\delta$ into a distribution of values. This effect is explored in Fig. 3c, which shows the probability distribution of closest approach distances of electrolyte atoms in short first-principles molecular dynamics simulations of a graphene-water interface at 0, –0.25, and +0.25 V, computed within the ESM framework (see Supplemental Material for details). The broadening in our calculated distributions is nonnegligible (width at half maximum ~1.0 Å); this will enhance the average capacitance because the sensitivity to $\delta$ is higher at shorter approach distances (Fig. 3b). (Note that the approach distance is based on the atomic positions rather than electron density, and hence is systematically larger than $\delta$.) The distributions also skew and shift with applied bias (by approximately +0.4 [-0.2] Å for positive [negative] bias), meaning the contact layer effect on capacitance should be asymmetric with voltage. We conclude that not only does the average electrolyte approach distance impact the interfacial capacitance of the electrode, but also the distance distribution.

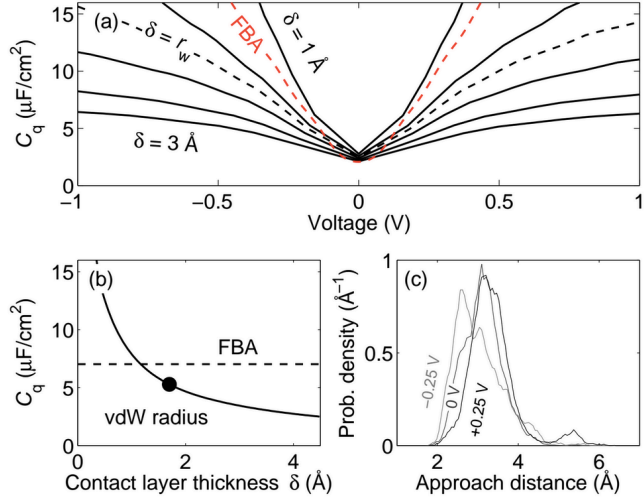

FIG. 3. *(a) Calculated differential quantum capacitance as a function of voltage for δ going from 1 to 3 Å in increments of 0.5 Å (solid lines); the dashed line shows the FBA-derived capacitance. (b) Calculated differential quantum capacitance for the two-sided charging of SLG at a bias of +0.2 V as a function of contact layer thickness δ. The maximum value of δ considered, 4.5 Å, corresponds to half of the simulation cell length. The black circle shows the capacitance for δ equal to the carbon van der Waals radius, and the horizontal dashed line shows the FBA-derived capacitance. (c) Average approach distance from a graphene electrode to the closest liquid atom in first-principles dynamics simulations of a water-graphene interface at 0 V, +0.25 V, and −0.25 V.*

In summary, we describe a first-principles method for simulating capacitive charging that allows for direct calculation of the quantum capacitance in the presence of the interfacial electric field induced by the double layer. We find that in graphene, space-charge effects lead to significant systematic errors in the computed capacitance if the field is neglected. The observed capacitance of the electrode also depends on the choice of electrolyte, which couples to the electrode capacitance through the interfacial contact layer. Our results suggest two avenues for improvement of the quantum capacitance in graphene-derived supercapacitors and similar electrochemical devices. The first is to address the screening within the electrode by increasing the thickness or modifying the dielectric properties [10]. The second is to engineer the electrolyte to tune its interaction with the electrode and modify the approach distance distribution. This could be done with the a proper choice of electrolyte or additive [32–34], or by leveraging confinement effects [35].

This work was performed under the auspices of the U.S. Department of Energy under by LLNL under Contract DE-AC52-07NA27344. Funding was provided by LDRD Program Grant 12-ERD-035, with computing support from the LLNL Institutional Computing Grand Challenge Program.